\documentclass[twocolumn,prb,showpacs,preprintnumbers,amsmath,amssymb]{revtex4}

\usepackage{natbib}

\usepackage{graphicx}
\usepackage{dcolumn}
\usepackage{bm}
\usepackage[T1]{fontenc}
\usepackage{multirow}
\usepackage{color}

\begin{document}

\title{Kagome Metal GdNb$_6$Sn$_6$: A 4$d$ Playground for Topological Magnetism and Electron Correlations}

\author{Yusen Xiao$^{1}$, Qingchen Duan$^{1}$, Zhaoyi Li$^{2,3}$, Shu Guo$^{2,3}$, Hengxin Tan$^{4,5}$, Ruidan Zhong$^{1,4}$\footnote{E-mail:rzhong@sjtu.edu.cn } }

\address
{  
	 $^{1}$    Tsung-Dao Lee Institute, Shanghai Jiao Tong University, Shanghai, 201210 China\\  $^{2}$ Shenzhen Institute for Quantum Science and Engineering, Southern University of Science and Technology, Shenzhen 518055, China\\
    $^{3}$ International Quantum Academy, Shenzhen 518048, China\\
      $^{4}$     School of Physics and Astronomy, Shanghai Jiao Tong University, Shanghai, 200240 China \\
       $^{5}$ Department of Condensed Matter Physics, Weizmann Institute of Science, Rehovot 7610001, Israel
	     }

\date{\today}

\begin{abstract}

Magnetic kagome metals have garnered considerable attention as an ideal platform for investigating intrinsic topological structures, frustrated magnetism, and electron correlation effects. In this work, we present the synthesis and detailed characterization of GdNb$_6$Sn$_6$, a metal that features a niobium-based kagome lattice and a frustrated triangular gadolinium network. The compound adopts the HfFe$_6$Ge$_6$-type crystal structure, with lattice parameters of $a$ = $b$ = 5.765(4) Å and $c$ = 9.536(8) Å. Magnetic susceptibility and specific heat measurements reveal a magnetic transition near 2.3 K. Electrical transport data confirm metallic behavior, unsaturated positive magnetoresistance, and a hole-dominated multiband Hall effect. Furthermore, first-principles calculations indicate that Nb-4$d$ orbitals predominantly contribute to the electronic states near the Fermi energy, with the band structure showing multiple topologically nontrivial crossings around the Fermi surface. This study also compares GdNb$_6$Sn$_6$ with GdV$_6$Sn$_6$, highlighting their similarities and differences. Our findings pave the way for exploring $R$Nb$_6$Sn$_6$ ($R$ = rare earth) with customized substitutions of $R$ sites to fine-tune their properties.

\end{abstract}

\pacs{XXX}


\maketitle
\section{\label{sec:level1}Introduction}

The kagome lattice, a two-dimensional network of corner-sharing triangles, has been a longstanding subject of interest in condensed matter physics due to its potential to host a wide array of exotic quantum states\cite{1,2,3}. Extensive research has revealed intriguing physical properties in kagome materials, such as frustrated magnetism, unconventional superconductivity, charge density waves, the quantum Hall effect, and quantum spin liquids\cite{4,5,6,7,8,9}.These phenomena imply the discovery of new kagome materials frequently heralds the manifestation of novel and intriguing physical phenomena.

Recently,  kagome layer ternary rare-earth compounds in the $R$$T$$_6$Sn$_6$ family (where $R$ represents rare-earth elements and $T$ denotes transition metals such as V or Mn) have attracted considerable attention\cite{Mn1,Mn2,Mnantif,Mnhall,Mnhall2,Mnmagspin,TbMn,TbMn2,GdV1,GdV2prb,GdVmag,RV,RV2,Sc2,ScV1,ScV3,YVos,TiV,TbV,TbVzeeman,ThV,V-arpes}. Among these, $R$Mn$_6$Sn$_6$ compounds stand out for their rich electronic and magnetic properties, including magnetic spin chirality, long-range double-cone spin structures, incommensurate antiferromagnetic arrangements, both anomalous and topological Hall effects\cite{Mn1,Mnmagspin,Mnhall,Mnhall2}. Interestingly, TbMn$_6$Sn$_6$ has been shown to host Chern-gapped Dirac fermions and exhibit topological charge-entropy scaling related to Berry curvature\cite{TbMn,TbMn2}. In the $R$V$_6$Sn$_6$ series, diverse quantum phenomena have been observed. For instance, quantum oscillations occur in YV$_6$Sn$_6$ and (Ti, Zr, Hf)V$_6$Sn$_6$, while ScV$_6$Sn$_6$ demonstrates charge density waves\cite{YVos,TiV,ScV1,Sc2,ScV3}. TbV$_6$Sn$_6$ also displays Spin Berry curvature-enhanced orbital Zeeman effect behavior\cite{TbVzeeman}. Additionally, nonlinear Hall effects are prevalent in these materials\cite{ThV,TiV,GdV2prb}. These findings demonstrate that the versatility of the rare-earth element $R$ allows for the tuning of physical properties, while the transition metal $T$ plays a key role in determining the defining characteristics of kagome materials. Therefore, the discovery and exploration of new transition metal kagome materials are of particular importance. In the past, research on 166-type compounds has predominantly focused on those containing 3$d$ transition metals, whereas studies on their 4$d$ counterparts remain limited. Early studies have identified YNb$_6$Sn$_6$ and TbNb$_6$Sn$_6$ as byproducts\cite{YNb,TbNb}, which suggests the existence of a niobium-based 166 family, providing a good reference for us to explore and study 4$d$ 166-type compounds. Noting that the $Ln$Nb$_6$Sn$_6$ ($Ln$: Ce-Lu, Y) series of compounds recently have been reported, revealing complex magnetic behaviors and density waves transition within these niobium-based materials. Although this material is made of 4$d$ transition-metal Nb sublattices, it is determined as non-magnetic\cite{LnNb}, similar to V-based 166 materials. This highlights the growing significance of advancing research on niobium-based kagome systems to explore their potential in condensed matter physics.

In this paper, we present a comprehensive study that focuses on the synthesis, structure, and physical properties of a Nb-based kagome metal, GdNb$_6$Sn$_6$. X-ray powder diffraction analysis confirms that the compound crystallizes in a HfFe$_6$Ge$_6$-type structure. Magnetic susceptibility combined with specific heat measurements indicates the formation of long-range magnetic ordering below 2.3 K. The electrical transport measurements confirm metallic behavior. Additionally, unsaturated positive magnetoresistance and multiband Hall effect at low temperatures suggest a complex transport mechanism. First-principles calculations reveal that the electronic structure of GdNb$_6$Sn$_6$ includes Dirac points and Van Hove singularities near the Fermi energy, originating from Nb-4$d$ electrons. Furthermore, we compare GdNb$_6$Sn$_6$ and GdV$_6$Sn$_6$, emphasizing both their similarities with distinctions. Our results suggest that niobium-based 166 kagome metals are an intriguing platform for studying the interplay between rare-earth layers and nonmagnetic kagome lattices.

\section{\label{sec:level2}Experimental}

The synthesis of GdNb$_6$Sn$_6$ involved the use of high-purity gadolinium(Gd)chips (99.9 Alfa), niobium(Nb) powders (99.95 Alfa), and tin(Sn) powders (99.999 Alfa). These components were homogeneously mixed in stoichiometric ratios using an agate mortar and subsequently cold-pressed into a pellet under an argon atmosphere within a glove box to preserve a controlled environment. The pellet underwent arc melting on a water-cooled copper hearth, with repeated melting cycles and flipping to ensure compositional uniformity. The consolidated button was then placed into an alumina crucible, vacuum-sealed within a silica ampoule, and annealed at 900°C for five days to enhance crystallinity.

For the cultivation of GdNb$_6$Sn$_6$ single crystals, the Sn flux method was utilized. Gd, Nb, and Sn were combined in a 1:6:40 ratio within an alumina crucible from Canfield Crucible Sets, and then sealed under vacuum in a silica ampoule. The sealed mixture was heated to 1000°C over a period of 3 hours, maintained at this temperature for 30 hours to achieve homogenization, and then cooled to 800°C at a controlled rate of 1°C/h. Post-cooling, the ampoule was rapidly centrifuged to separate the excess Sn flux. The crystal surfaces were treated with a dilute HCl solution to eliminate any residual tin. The resulting platelet-shaped single crystals, characterized by a pristine hexagonal ab plane as the basal facet and measuring approximately 1×0.5×0.2 mm$^3$, are depicted in the inset of Figure 1b.

XRD analysis was performed using a Bruker D8 AdvanceEco diffractometer, collecting data from 10°-90° in 2$\theta$ and refining the data via the Rietveld method with GSASII software. Electrical resistivity and magnetoresistance measurements were conducted using a Quantum Design PPMS, employing the standard four-probe method with current flowing within the $a$$b$ plane. The Hall coefficient was determined by sweeping the magnetic field from positive to negative at a fixed temperature, with the magnetoresistance component subtracted. The transverse resistivity $\rho$$_x$$_y$ (Figure 6b) showed noise, potentially due to the sample's low resistivity. DC magnetization measurements down to 1.8 K were executed on a VSM integrated with a PPMS. Specific heat capacity measurements were also carried out down to 1.8 K using a relaxation technique on the PPMS.

First-principles calculations were conducted with the plane-wave-based Vienna ab initio simulation package (VASP)\cite{cal1}, employing the projector augmented wave (PAW) method\cite{cal2}. The exchange-correlation interaction between electrons is mimicked by the generalized gradient approximation (GGA) as parametrized by Perdew, Burke, and Ernzerhof\cite{cal3}. The plane-wave basis set cutoff energy of 250 eV was used, along with a 16×16×8 Gamma-centered Monkhorst-Pack grid for adequate Brillouin zone sampling. The crystal structure is optimized with a force convergence threshold of 1 meV/Å per atom. The optimized lattice constants are 5.805 Å and 9.628 Å for the in-plane and out-of-plane ones, respectively. The density of states (DOS) is calculated using the tetrahedron method. In all calculations, the PAW potential of Gd$^{3+}$ is employed where $f$ electrons are taken as the core state.

\section{\label{sec:level3}Results and Discussion}

\begin{figure}[tbp]
\includegraphics[width=7cm]{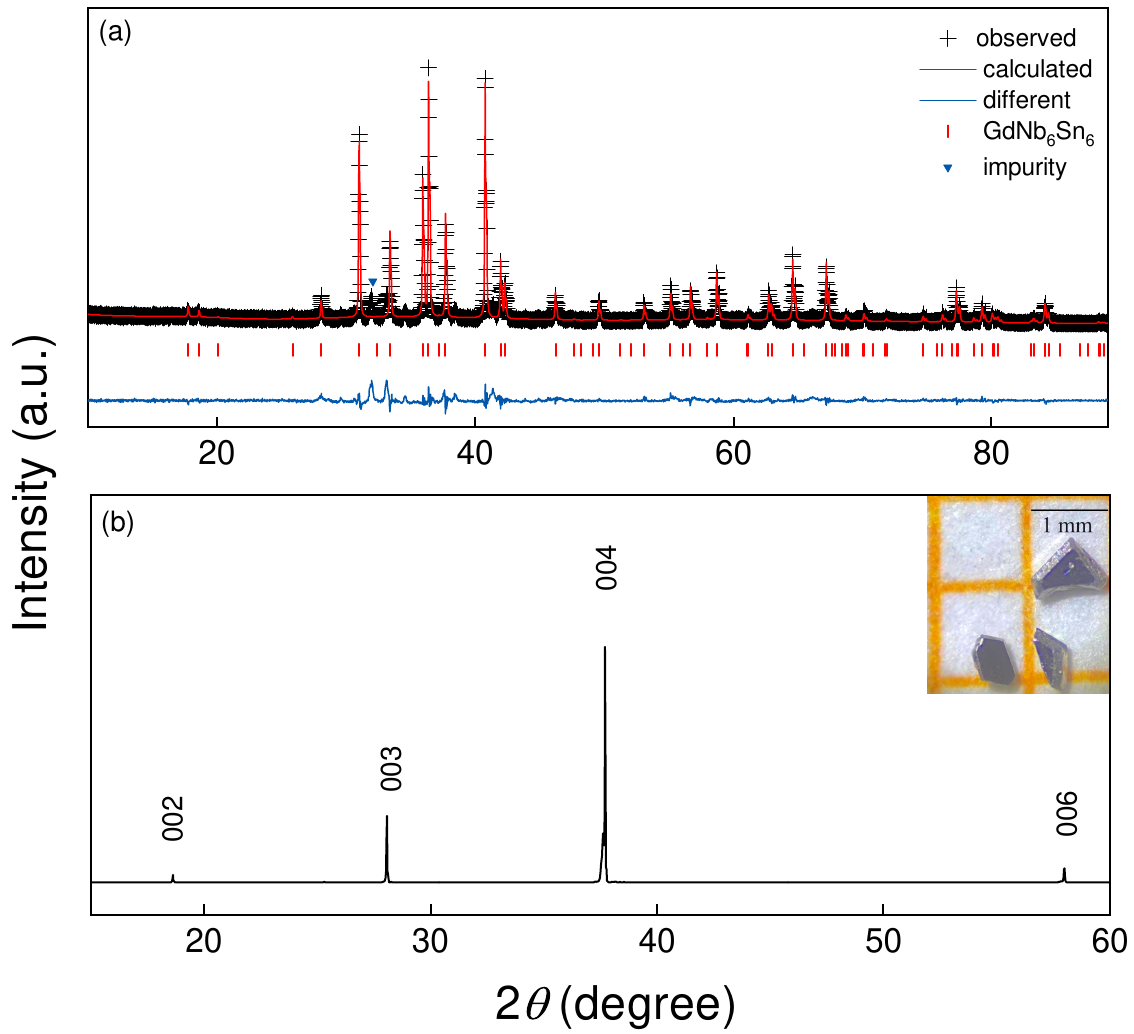}
\caption {(Color online)  (a) Rietveld refinement profile of GdNb$_6$Sn$_6$ polycrystalline sample. (b) X-ray diffraction pattern of the GdNb$_6$Sn$_6$ single crystal.}
\label{fig1}
\end{figure}

\begin{figure*}[tbp]
	\includegraphics[width=12cm]{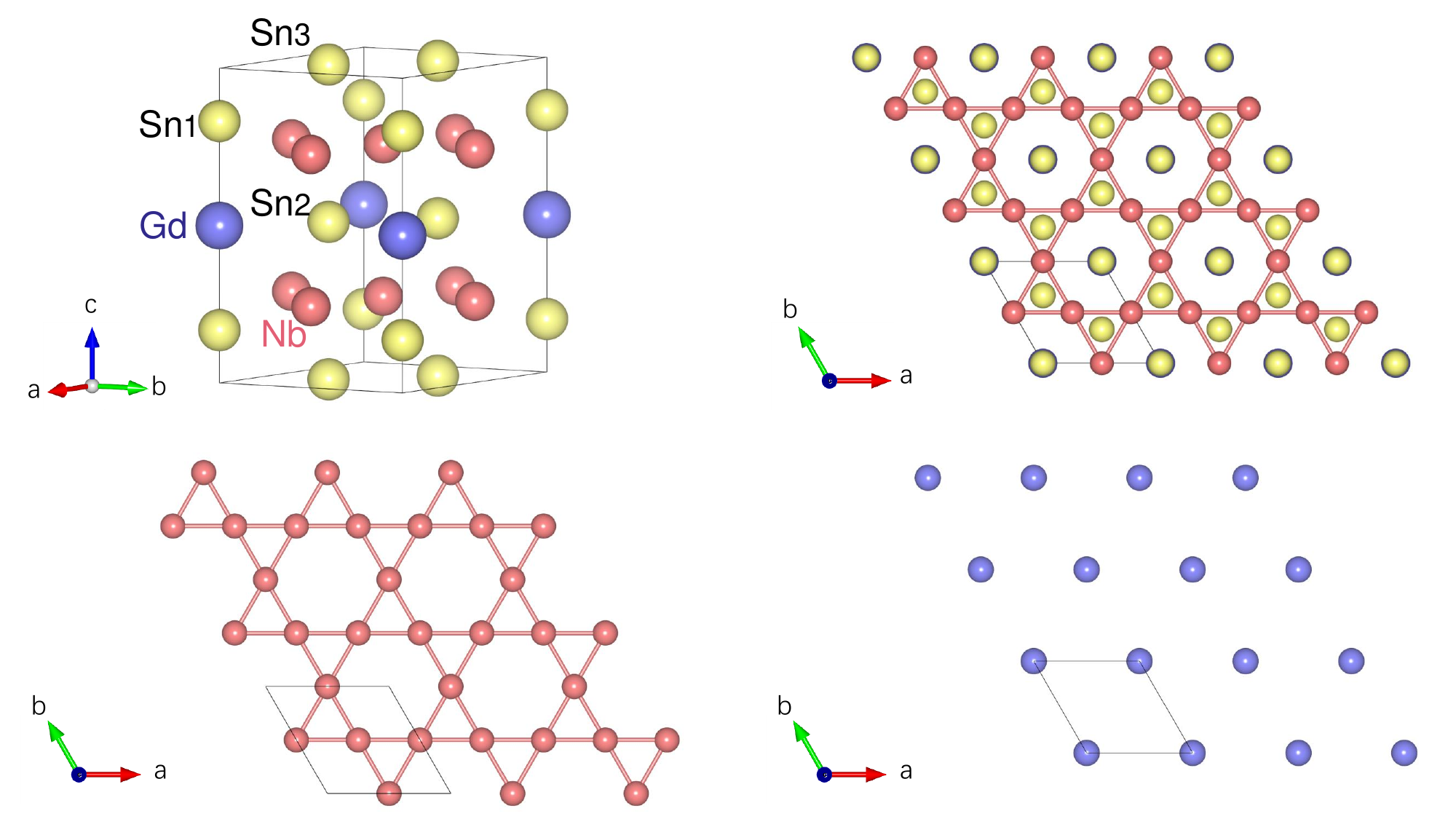}
	\caption {(Color online) (a) The crystal structure of GdNb$_6$Sn$_6$. The various tin positions are denoted as Sn1, Sn2, and Sn3. (b) A top-down view of the crystal structure along the $c$-axis. (c) A two-dimensional kagome network formed by niobium atoms. (d) A triangular lattice of Gd ions interlaced between the kagome planes, as viewed along the $c$-axis.}
	\label{fig2}
\end{figure*}

 Figure \ref{fig1}a presents the XRD pattern for GdNb$_6$Sn$_6$, accompanied by the Rietveld refinement profile characteristic of the HfFe$_6$Ge$_6$-type structure, which belongs to the space group $P6/mmm$. Figure 1b illustrates the XRD pattern of a single-crystalline sample. The refined structural parameters are summarized in Table I. The lattice parameters derived from the refinement are a = 5.765(4)Å and c = 9.536(8)Å. GdNb$_6$Sn$_6$ shares a structural analogy with $R$V$_6$Sn$_6$\cite{RV,RV2}, featuring a stacking sequence of Nb-ion kagome layers along the crystallographic $c$-axis. The Nb-Nb distance within the kagome layer is 2.8827 Å, exceeding the V-V bond length observed in the related compound $R$V$_6$Sn$_6$\cite{GdV2prb,ThV}. Interestingly, both GdNb$_6$Sn$_6$ and GdV$_6$Sn$_6$ exhibit a very similar $c$/$a$ ratio, approximately 1.65, suggesting that elemental substitution results in isotropic three-dimensional changes to the unit cell.
 
\begin{table}[!h]
\caption{Crystallographic data of GdNb$_6$Sn$_6$ at room temperature.} \label{tab1}
\setlength{\tabcolsep}{0.8mm}
\renewcommand\arraystretch{1.2}
\begin{center}
\begin{tabular}{ccccccc}
\hline
&Compounds       &&& GdNb$_6$Sn$_6$ &   \\
&space group     &&& $P6/mmm$          & \\
&$a$ (\AA)     &&& 5.765(4)      & \\
&$c$ (\AA)     &&& 9.536(8)       & \\
&$V$ (\AA$^3$) &&& 274.54        & \\
&Nb-Nb distance (\AA) &&& 2.8827        & \\
&$R_{\rm{wp}}$ (\%)   &&& 9.76           & \\
&$\chi^2$             &&& 1.85           & \\
&$Z$             &&& 1               & \\
\hline
atom  &  site    &  $x$  &  $y$  &  $z$             & Occ. \\
\hline
Gd   &  1b      &  0    &   0   &   0.5      &  1&          \\
Nb    &  6i      &  0.5   &  0.5  & 0.7421     &  1   &          \\
Sn1    &  2e      &  0  &  0  & 0.8320   &  1     &       \\
Sn2     &  2d      &  0.3333    &   0.6667   &  0.5     &  1   &          \\
Sn3     &  2c      &  0.3333    &   0.6667   &  0     &  1    &          \\
\hline
\hline
\end{tabular}
\end{center}
\end{table}

The Gd and Nb ions occupy specific crystallographic sites, while Sn ions are distributed across three distinct crystallographic positions, denoted as Sn1, Sn2, and Sn3, as outlined in Figure  \ref{fig2}. The unit cell is composed of alternating layers of Nb$_3$Sn1, separated by two distinct layers of Sn3 and GdSn2, thus forming a layered structure of Nb$_3$Sn1, GdSn2, Nb$_3$Sn1, Sn3 along the $c$-axis. Figure \ref{fig2}(b) offers a top-down view of the crystal structure, highlighting the kagome arrangement of Nb atoms within the $ab$ plane. The Sn2 and Sn3 sites are positioned to form stannene planes interspersed between the Nb kagome layers. The isolated Nb kagome network is depicted in Figure  \ref{fig2}(c). The interstitial Gd atoms arrange themselves into a triangular lattice plane, as shown in Figure  \ref{fig2}(d). It is worth noting that the triangular lattice may introduce additional geometric frustration and electronic complexity to the already intricate kagome network.

\begin{figure}[tbp]
\includegraphics[width=8.5cm]{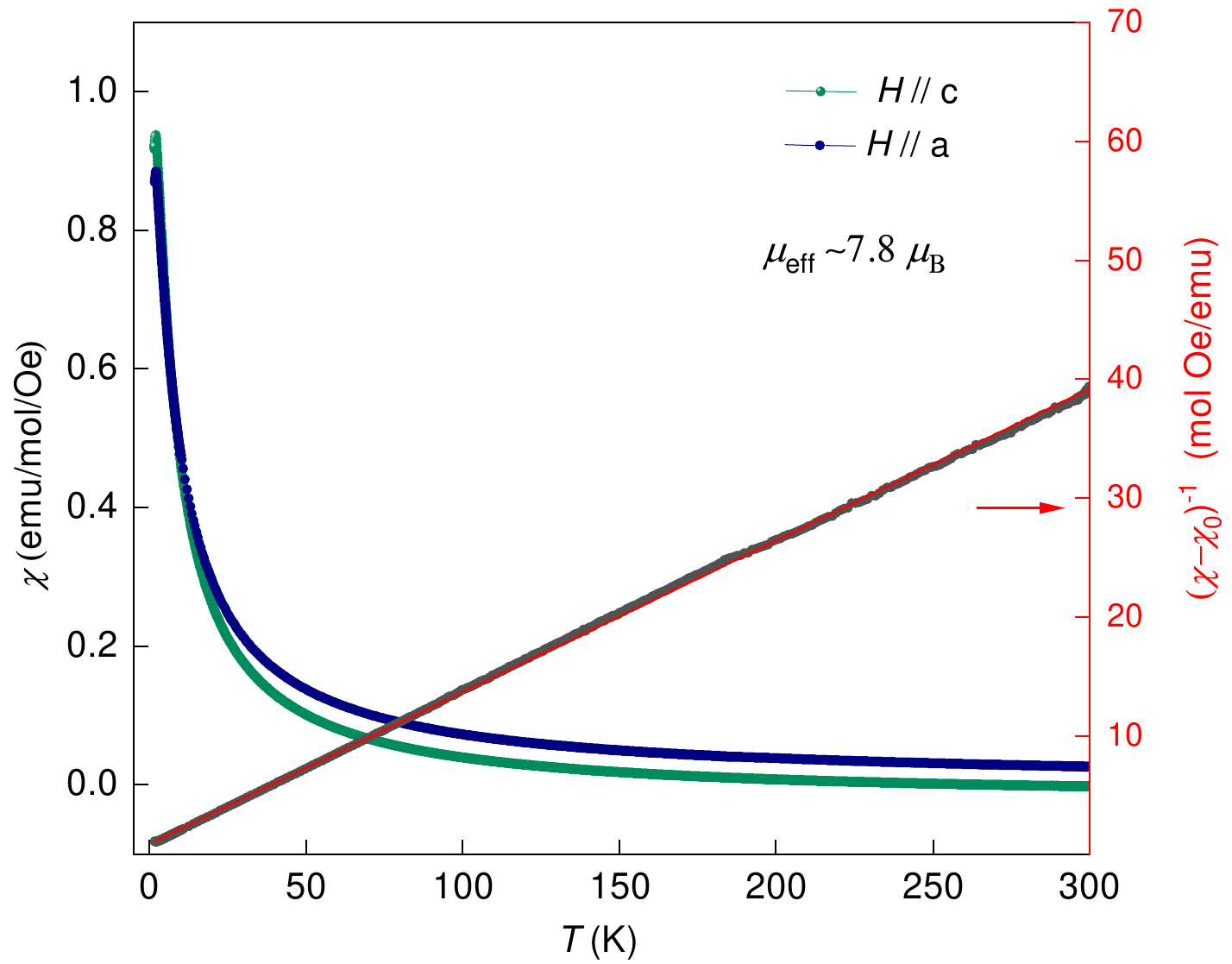}
\caption {(Color online) Temperature-dependent magnetization (ZFC) data for GdNb$_6$Sn$_6$ with fields of  $\mu_0$$H$ = 1 T oriented both parallel and perpendicular to the $c$-axis. black line: Temperature-dependent inverse susceptibility, ($\chi$-$\chi$$_0$)$^{-1}$, for fields aligned parallel to the $a$-axis. The red solid line represents a Curie-Weiss fit to the data (see the text in details).  }
\label{fig3}
\end{figure}

\begin{figure}[tbp]
\includegraphics[width=8.5cm]{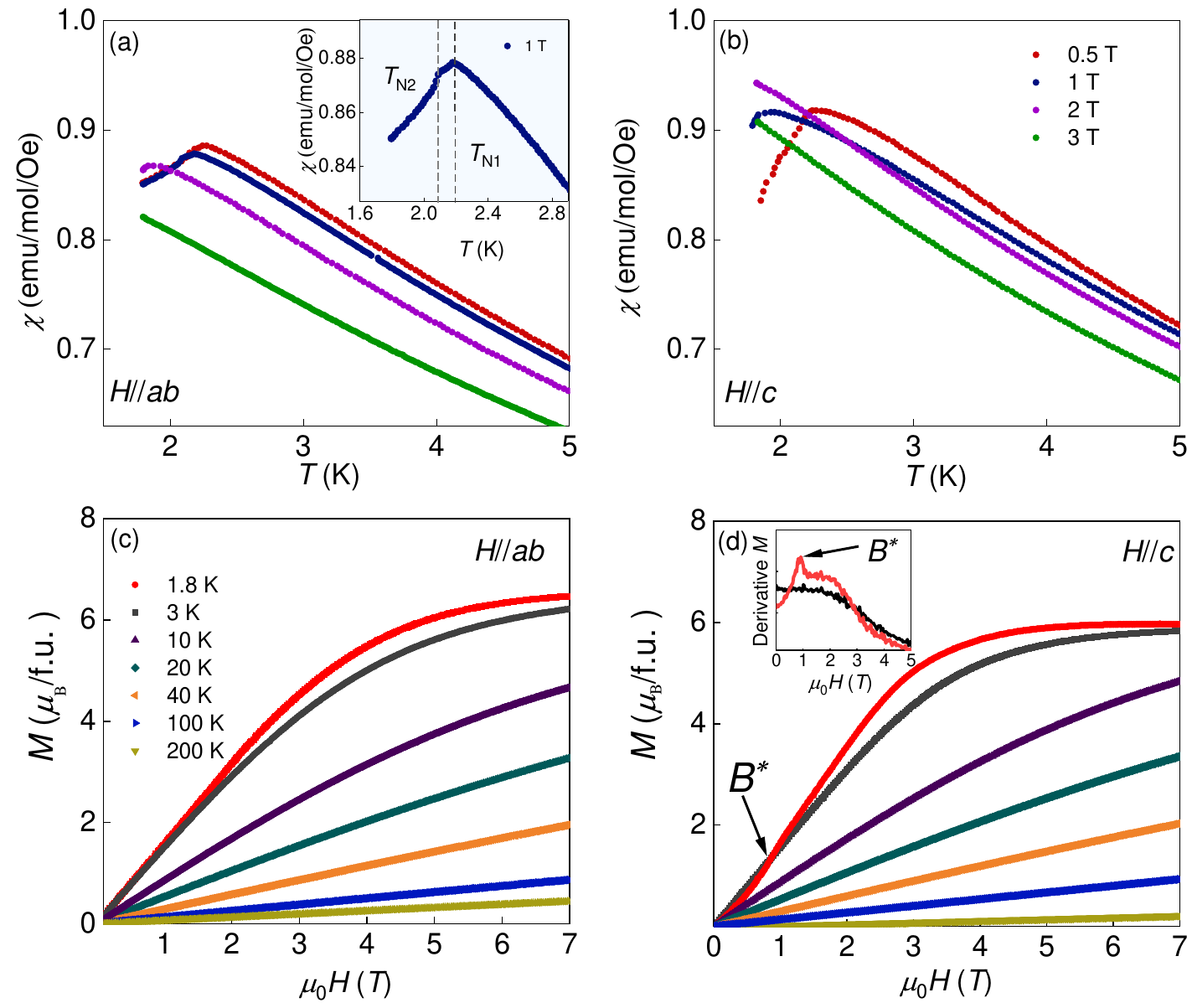}
\caption {(Color online) (a) and (b) Temperature dependence of magnetic susceptibility for GdNb$_6$Sn$_6$ sample under different external magnetic fields.The illustration zooms in on the magnetic susceptibility data under 1 T, allowing for a clear observation of two magnetic phase transitions, $T$$_{N1}$ and $T$$_{N2}$. (c) and (d) Field dependence of magnetization for GdNb$_6$Sn$_6$ at various temperatures. The derivative of the 1.8 K data is shown in the inset.}
\label{fig4}
\end{figure}

Figure \ref{fig3}(a) presents the temperature dependence of the magnetic susceptibility $\chi$($T$) under an external field of 1 T for the single-crystalline sample of GdNb$_6$Sn$_6$. Measurements were conducted using both zero-field-cooling (ZFC) and field-cooling (FC) protocols. As the ZFC and FC data align almost perfectly, only the ZFC results are displayed. Comparing the magnetic susceptibility in different magnetic field directions, a slight anisotropy is observed, indicative of minor planar anisotropy, similar to that observed in GdV$_6$Sn$_6$\cite{GdV1,GdV2prb}. The magnetic susceptibility increases gradually with decreasing temperature but rises sharply below 50 K, reaching a peak at approximately 2.3 K. Below this temperature, the magnetic susceptibility declines, demonstrating antiferromagnetic behavior as a whole (Further details are discussed in Figure \ref{fig4}). In addition, the $\chi$($T$) data can be modeled using an extended Curie-Weiss formula,
\begin{center}
$\chi=\frac MH=\chi_0+\frac C{T-\theta}$
\end{center}
where $\chi$$_0$, $\theta$ and $C$ are the fitting parameters. $C$ is defined as $\mu_0\mu_{\mathrm{eff}}^2/3k_\text{B}$, with $\mu_{\mathrm{eff}}$ representing the local magnetic moment at the Gd site. The formula can also be expressed in a linear form: ($\chi$-$\chi$$_0$)$^{-1}$ =($T$- $\theta$)/C. The slope of ($\chi$-$\chi$$_0$)$^{-1}$ versus $T$ provides the value of 1/C, while the intercept gives $\theta$/C. Black line illustrates the temperature dependence of ($\chi$-$\chi$$_0$)$^{-1}$  for GdNb$_6$Sn$_6$, yielding $\chi$$_0$ = 1.41× 10$^{-4}$emu mol$^{-1}$, $C$  = 7.69 emu K mol$^{-1}$, and $\theta$= -6.1 K. Negative $\theta$ indicates the net magnetic interaction is dominated by antiferromagnetic correlations. The Curie-Weiss constant $C$, corresponding to an effective moment $\mu$eff=7.84$\mu$$_\text{B}$/f.u., agrees well with the expected value for Gd$^{3+}$ ions with $J$ = 7/2 and the low-lying crystal field multiplets that are populated above 100 K.

Geometric frustration or competing interactions on a triangular lattice may result in complex magnetic orders. Figures \ref{fig4}(a) and (b) illustrate the temperature dependence of magnetic susceptibility under varying magnetic fields. The low-temperature decrease is significantly suppressed with increasing field strength, particularly in the $H$//$c$ orientation, where at 2 T, the $T$$_N$ is below 1.8 K. As shown in the inset of Figure \ref{fig4}a, two distinct phase transitions, $T$$_{N1}$ and a minor $T$$_{N2}$, are discernible for $H$//$a$. This phenomenon is also observed in the isostructural compound GdV$_6$Sn$_6$\cite{GdV2prb,GdVmag}, suggesting the potential for non-collinear or spiral magnetic ordering.

Figures \ref{fig4}(c) and (d) demonstrate the  field dependence of magnetization at different temperatures for GdNb$_6$Sn$_6$. No hysteresis is observed at low temperature. When the field is aligned parallel to the crystal surface, the saturation magnetization reaches approximately 6.5 $\mu$$_\text{B}$, close to the expected value of 7 $\mu$$_\text{B}$ for free Gd-ion. In contrast, the magnetization is more responsive to fields parallel to the $c$-axis, approaching saturation at around 4 T. In particular, a metamagnetic transition is found at 1.8 K, with a critical field $B^*$ around 1 T according to the derivative analysis.

\begin{figure}[tbp]
\includegraphics[width=7cm]{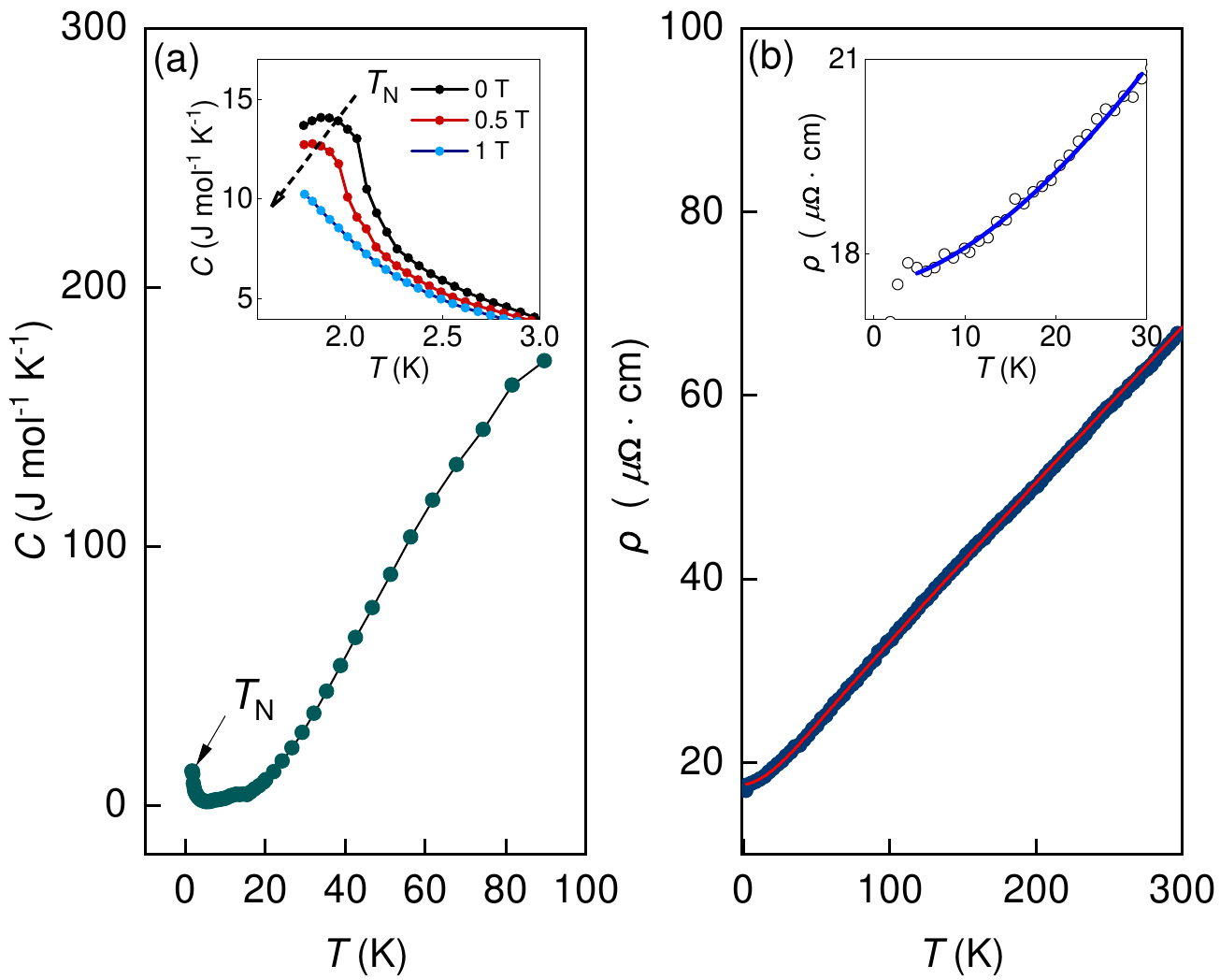}
\caption {(Color online) (a) Temperature-dependent specific heat for GdNb$_6$Sn$_6$. The inset gives the  the specific heat under different magnetic field. (b) Temperature dependence of the electrical resistivity for the GdNb$_6$Sn$_6$, which basically obeys an extended Bloch-Grüneisen  formula, as detailed in the accompanying text. The inset shows a power-law dependence in the low-temperature regime. }
\label{fig5}
\end{figure}

The temperature dependence of the specific heat is depicted in Figure \ref{fig5}(a).   The specific heat increases rapidly at low temperatures, with an inflection point at about 2 K, and a full $\lambda$ shape requires lower specific heat data.   The inset of Figure \ref{fig5}(a) shows the specific heat at low temperature under different magnetic fields.   As the magnetic field increases, the phase transition is suppressed to lower temperatures, which is consistent with the magnetic susceptibility measurement.
The temperature dependence of resistivity, denoted as $\rho(T)$, for the GdNb$_6$Sn$_6$ single crystal sample is presented in Figure \ref{fig5}(b). At 300 K, the zero-field resistivity is measured to be 70 $\mu \Omega$ cm, which is comparable to that of GdV$_6$Sn$_6$. As temperature decreases, resistivity also diminishes, indicating a metallic behavior. Due to the low phase transition temperature and the minimal presence of the superconducting Sn in the sample, the impact of magnetic ordering on resistance is undetectable. However, resistivity data above 4 K could be well modeled using the extended  Bloch-Grüneisen formula\cite{BB1,BB2}, with $n$ as the variable.

\begin{center}
	$\rho(T)=\rho_0+A\left(\frac{T}{\Theta_R}\right)^n\int_0^{\frac{\Theta_R}{T}}\frac{\chi^n}{(\mathrm{e}^x-1)(1-\mathrm{e}^{-x})}\mathrm{d}x$
\end{center}
The fitted parameters are   $\rho$$_0$= 0.036 m$\Omega$ cm, A = 0.133 m$\Omega$ cm, $\Theta$$_{R}$=300, and n=1.7. Note that the value of $n$ matches the power-law function fitted with the power exponent $\alpha$=1.67, using the resistivity data in the low-temperature range of 4 K < T < 30 K, expressed as $\rho(T)=\rho_0+BT^\alpha$. Based on the values of $n$ or $\alpha$, GdNb$_6$Sn$_6$  deviates from the  Fermi liquid behavior, which implies the possibility of strong correlations. 

\begin{figure}[tbp]
	\includegraphics[width=7cm]{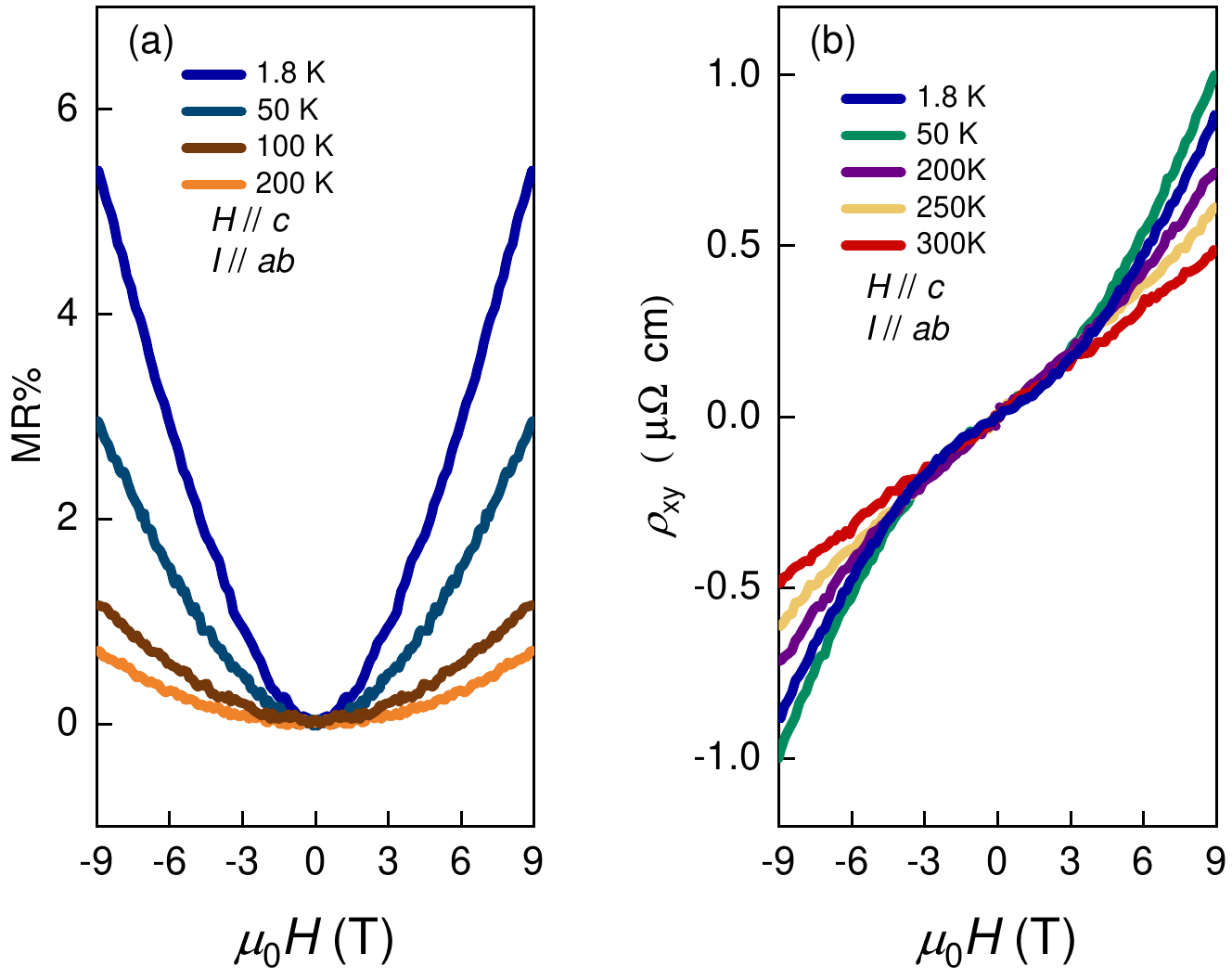}
	\caption {(Color online) ) The variation of magnetoresistance (MR) 
with temperature, defined as $[\rho_{xx}(H)-\rho_{xx}(H=0)]/\rho_{xx}(H=0)\times100\%$. (b) Hall resistivity $\rho_{xy}$ measured at different  temperatures. To eliminate the influence of longitudinal resistivity, $\rho_{xy}$ is derived as the average difference between positive and negative magnetic field measurements, calculated as $\rho_{xy}=[\rho_{xy}(+H)-\rho_{xy}(-H)]/2$.}
	\label{fig6}
\end{figure}

To further explore the magnetoresistance (MR) effect, we performed supplementary measurements of the longitudinal resistivity's response to varying magnetic fields at different temperatures, as shown in Figure \ref{fig6}(a). The observed data displays a minor asymmetry when the magnetic field direction is reversed, attributed to the superposition of the MR effect with a Hall effect. As a result, the extraction of the MR signal is accomplished by calculating the mean resistivity values across opposing magnetic field polarities. The current is directed along the ${ab}$-plane, while the magnetic field is aligned along the $c$-axis. We observe a moderate MR effect across the entire temperature range, with no indication of saturation even at a magnetic field strength of 9 T. However, no negative MR was observed despite the temperature drop to 1.8 K. Hall resistance ($\rho_{xy}$) measurements were performed over a temperature range from 1.8 K to 300 K, with the magnetic field oriented parallel to the $c$-axis and the current flowing within the ${ab}$-plane, as depicted in Figure \ref{fig6}(b). The results reveal a weak temperature dependence of $\rho_{xy}$, yet a nonlinear decrease is observed as the temperature drops below 250 K, indicative of multiband transport. Above 250 K, the $\rho_{xy}$ curves exhibit linearity. Interestingly, the positive sign of $\rho_{xy}$ across the 1.8-300 K range suggests a predominance of hole carriers, while the sister compound $R$V$_6$Sn$_6$ typically exhibits electron carriers. Note that the nonlinearity in Hall resistivity at low temperatures is characteristic of the  $R$166 family\cite{GdV2prb,TbV,ThV,TiV}.

\begin{figure}[tbp]
	\includegraphics[width=8cm]{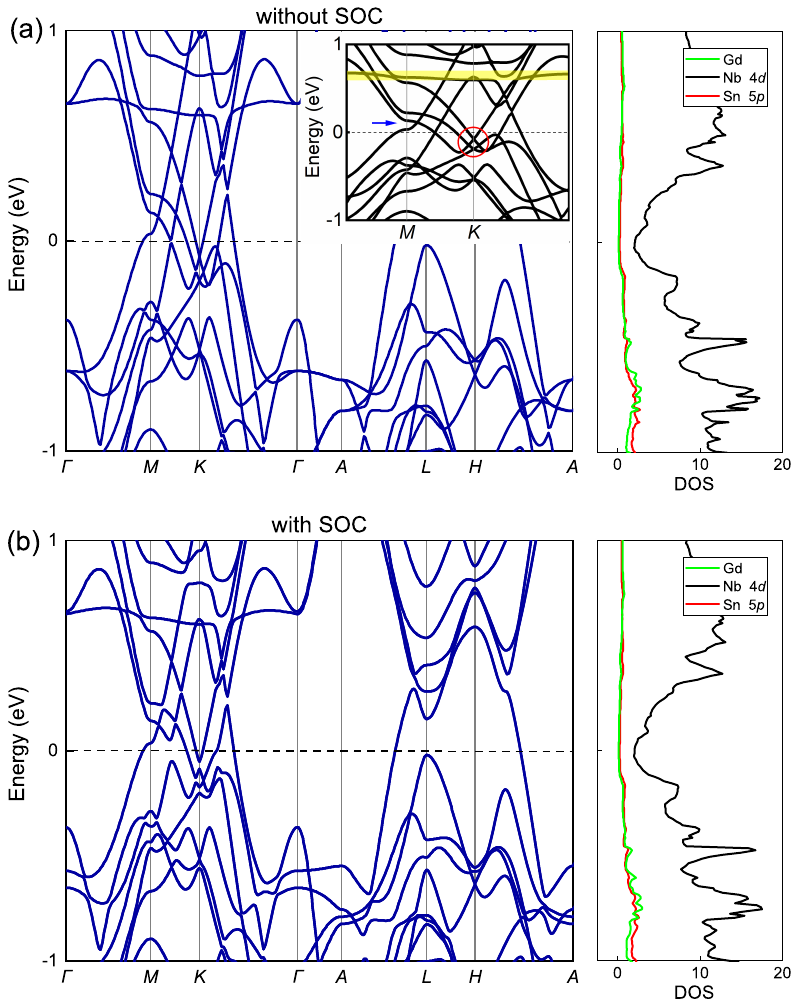}
	\caption {(Color online) First-principles electronic structure calculations of GdNb$_6$Sn$_6$. (a) without SOC (b) With SOC. }
	\label{fig7}
\end{figure}

Electronic structure information is a crucial pathway to understanding the properties of kagome materials.  The calculated electronic band structure and density of states (DOS) are shown in Figure \ref{fig7}, which indicates the metalicity of GdNb$_6$Sn$_6$.  In the low-energy range, the Nb $d$ orbitals dominate the band structure.  The overall band structure of GdNb$_6$Sn$_6$ is similar to the $R$V$_6$Sn$_6$ series, which exhibits typical kagome band features, including the Dirac band crossings at the Brillouin zone (BZ) corner point $K$, van Hove singularities (vHSs) at the BZ edge point $M$, and flat bands on the $\Gamma$-$M$-$K$ plane.  It has been reported that the Dirac cones of GdV$_6$Sn$_6$ are located at the Fermi level \cite{GdV2prb}.  The heavier Nb atom substitution of V moves the Dirac point away from the Fermi energy (in GdNb$_6$Sn$_6$, the closest Dirac point is approximately 77 meV below the Fermi level).  After including the spin-orbit coupling (SOC), the Dirac points open gaps, which were widely observed in other kagome systems.  Dirac points close to the Fermi level promise the topological surface states in these kagome materials, although it might be challenging to detect them due to multiple bulk states at the same energy \cite{V-arpes,Tan2}.  The other notable feature is the presence of a vHS about 33 meV above the Fermi energy.  Such a close vicinity of the VHS might contribute to potential exotic properties, which had been widely discussed in the known kagome superconductor $A$V$_3$Sb$_5$\cite{kang2022twofold,hu2022rich}, kagome metal ScV$_6$Sn$_6$\cite{jiang2024van}, and the Co-based 166 kagome materials\cite{Co}.

On the other hand, the phonon dispersion of GdNb$_6$Sn$_6$ (see supplementary materials S1)exhibits no imaginary frequency modes, confirming its dynamic stability.  This observation aligns with our experimental results, which show no evidence of a density wave transition.  In contrast, the analogous compound LuNb$_6$Sn$_6$ has recently been reported to undergo a density-wave-like transition at 68 K\cite{LnNb}, highlighting the crucial role that rare-earth ($R$) elements play in driving structural instabilities within this material series.  Therefore, the successful synthesis of GdNb$_6$Sn$_6$ will facilitate comparative studies of these kagome systems\cite{feng2024catalogue}.

To date, vanadium-based 166 family compounds have been extensively studied, while niobium-based 166 compounds have long been neglected until the recent reports on the $Ln$Nb$_6$Sn$_6$ series by Ortiz et al\cite{LnNb}. Given that the transition metals V and Nb on the kagome layer are in the same main group, a reasonable comparison is warranted.

Both compounds derive their magnetism from gadolinium on the honeycomb lattice, with GdNb$_6$Sn$_6$ exhibiting a lower magnetic ordering temperature. Generally, when rare-earth ions are the sole source of magnetism in intermetallic compounds, magnetic ordering can be explained by the indirect Ruderman-Kittel-Kasuya-Yosida (RKKY) exchange interaction\cite{RKKY}, the strength of which varies with interatomic distances. Upon Nb substitution for V, the unit cell expands, modulating the RKKY interaction and thus affecting the magnetic ordering temperature. On the other hand, previous ARPES combined with density functional theory calculations have revealed significant hybridization between V 3$d$ and Sn 5$p$ electrons in GdV$_6$Sn$_6$, which is crucial for magnetic ordering at low temperatures\cite{GdVorbital}. In contrast, no such significant hybridization is observed in the DOS diagram of GdNb$_6$Sn$_6$, which may also influence magnetic ordering (see supplementary materials S2). Additionally, GdNb$_6$Sn$_6$ undergoes two magnetic phase transitions at low temperatures, both of which are rapidly quenched with increasing magnetic fields. This, along with the absence of low-field hysteresis in the magnetization data, suggests the formation of a weak, non-collinear magnetic order below 2.3 K, although this requires future confirmation through magnetic scattering measurements (neutron or resonant X-ray). Note that similar behavior is observed in GdV$_6$Sn$_6$\cite{GdV2prb}.

Although both GdNb$_6$Sn$_6$ and $R$V$_6$Sn$_6$ exhibit metallic behavior and nonlinear Hall effects, they show distinct dominant carrier types at low temperatures: GdNb$_6$Sn$_6$ is hole-dominated, while GdV$_6$Sn$_6$ is electron-dominated. Furthermore, no significant negative magnetoresistance is detected in GdNb$_6$Sn$_6$ at 1.8 K, unlike the prominent negative magnetoresistance observed in GdV$_6$Sn$_6$\cite{GdV2prb}. Combining MR and Hall effects, we speculate that the 166 family possesses complex transport mechanisms. Future measurements of magnetic transport properties at higher fields and lower temperatures, utilizing ARPES to observe band structure, are necessary.

\section{\label{sec:level4} CONCLUSIONS}

In conclusion, we successfully synthesized single crystals of the kagome metal GdNb$_6$Sn$_6$ and carried out a comprehensive study of their crystal structure, magnetic behavior, specific heat, and electrical transport properties. This material is characterized by a Nb-based kagome lattice coupled with a frustrated triangular Gd network. The single crystals display magnetic ordering below 2.3 K, as well as unsaturated magnetoresistance and a multiband Hall effect at low temperatures. Additionally, first-principles calculations reveal a complex electronic band structure with topological characteristics predominantly governed by Nb orbitals near the Fermi energy.

Building on the extensively studied and widely discussed 166 kagome materials based on 3$d$ transition metals (such as Mn and V), this work identifies and thoroughly characterizes a new class of 166 materials based on 4$d$ transition metals, specifically Nb, as discussed here. Our findings not only deepen the understanding of Nb-based kagome materials but also expand the 166 material family. Moreover, they shed light on the intricate interplay among topology, magnetism, and electron correlations. This work advances the study of 166-type compounds and underscores their potential as a versatile platform for investigating novel quantum phenomena.

\begin{center}
\textbf{ACKNOWLEDGEMENTS}
\end{center}
The work was supported by the National Natural Science Foundation of China with Grants No. 12374148, and 12334008, the the Ministry of Science and Technology of China under 2022YFA1402702, and the project by the National Key R$\&$D of China under 2021YFA1401600. Z.L. and S.G. acknowledge the financial support from the National Natural Science Foundation of China (22205091)

\bibliography{GdNb6Sn6}

\end{document}